\documentclass[pra,twocolumn,showpacs,amsmath,amssymb,floatfix,nobibnotes]{revtex4-2}

\usepackage{comment}
\usepackage[breaklinks]{hyperref}
\usepackage{xcolor}
\usepackage{soul}
\usepackage{dsfont}
\usepackage[T1]{fontenc}
\usepackage{physics}
\usepackage{multirow}
\usepackage{tikz}

\newcommand{\im}{\text{i}}

\def\12{\frac{1}{2}}

\begin{document}
	\title{Phase-preserving steady-state operations beyond Lindblad master equations}
	
	\author{Man Yin Cheung}
	\author{Mona Berciu}
	\author{Kyle Monkman}
	\email{Corresponding author: kyle.monkman@ucalgary.ca}
	\affiliation{Quantum Matter Institute, University of British Columbia, Vancouver, British Columbia V6T 1Z4, Canada}
	\date{\today}
	\begin{abstract}
		We consider a class of nonunitary operations that is naturally implemented via a combination of projective measurement and unitary operators. We implement these operations without using measurements by coupling the system to an infinite set of ancilla states and time-evolving with a single time-independent Hamiltonian. The infinite ancilla enables the main system to reach a local steady state that preserves initial phase information. We prove that these steady-state operations are not describable as a mapping from an initial state to a steady state of a time-independent Lindblad master equation acting only on the main system. As an additional degree of control, we show that the spectral resonance between the system and the ancilla acts as a switch that turns the operation on and off, and we derive a closed-form expression that quantifies the sharpness of this switch. We further find numerically that the phase preservation survives moderate disorder, over a time window set by the size of the ancilla. This Hamiltonian framework therefore provides a route to autonomous quantum control using the tunability of a closed system.

	\end{abstract}
	\maketitle
	\section{Introduction}
	Nonunitary operations are a central aspect of quantum control. They encompass several important protocols such as state preparation, qubit reset and quantum error correction. Since all closed time evolution is unitary, some type of external control mechanism is required to implement these operations. Broadly speaking, one can use projective measurements, coupling to a controlled ancilla state or coupling to a truly external environment. 
	
	Autonomous control refers to the implementation of such nonunitary operations without measurement or classical feedback \cite{Review,RefResSC,onDemandDissipation,onDemandDissipation2,reviewEngineerDissipation,autoWork,autonomousCooling,autonomousCooling2,autonomousMahler,autonomousERROR3,autonomousERROR5,autonomousERROR6,autonomousERROR7,errorCorrector,errorCorrector2,errorCorrector3,AutonomousGates}. This type of control is of recent interest as a way to bypass the decoherence inherently caused through a measurement process \cite{errorCorrector,errorCorrector3,AutonomousGates,reset1}. In particular, for quantum error correction, reducing measurements may be of importance for maintaining coherence. Currently, the main avenue for autonomous quantum error correction is to utilize coupling to an external bath where the correction process utilizes a Lindblad dissipator \cite{errorCorrector,errorCorrector2,errorCorrector3,autonomousERROR3,autonomousERROR5,autonomousERROR6,autonomousERROR7}. The key limitation of this direction is the limited control of a truly external bath, which has led to challenges in maintaining the phase information of the initial state.
	
	In principle, a Lindblad master equation can reach steady states which preserve information of the initial state \cite{AlbertJiang}. This requires that there are at least two steady states in the system that form a decoherence-free subspace. However, Lindbladians with multiple steady
	states still rely on highly fine-tuned coupling to an external
	bath. Furthermore, a general nonunitary operation cannot be implemented as a steady state of a time-independent Lindblad equation acting \textit{only} on the system. Instead, it fundamentally requires an expanded Hilbert space to map states from outside the decoherence-free subspace to within it and using a subsequent partial trace operation \cite{AlbertJiang2}.

	An alternative method for implementing nonunitary operations is through coupling the system to a controllable ancilla state \cite{autonomousCooling,ResetBound1}. The main advantage of this methodology is the precise tunability of the closed system+ancilla system, unlike a truly external bath. This tunability has promise for preserving phase information and implementing a broad class of nonunitary operations. 
	
	In this article, we implement a class of phase-preserving nonunitary operations by coupling a system to an infinite-dimensional ancilla. Although the full system+ancilla state is constantly time evolving, the main system approaches a well-defined steady state.
	Crucially, the implementation utilizes only a single time-independent Hamiltonian for the combined system and ancilla, thereby bypassing the need for control of an external bath. We prove that this class of operation is not implementable as a mapping from an initial state to a steady state of a time-independent Lindblad equation acting only on the main system.  
	
	We show that these operations are controllable via a  `spectral switch' mechanism where the energy levels of the system  select the states for operation. Once the coupling to the ancilla is turned on, the nonunitary operation is executed on states whose energy falls within a fixed window and suppressed outside of this window. The sharpness of this energy cutoff is characterized below. 
	
	\section{Class of nonunitary operations}
	\color{black}
	To formalize this, consider a Hilbert space $\mathcal{H_A}$ of finite dimension $N$ with orthonormal basis states $|i\rangle$. We define projectors $P=\sum_{i \in S_P} |i\rangle \langle i|$ and $Q=1-P$ where $S_P$ represents a subset of states. Let $U_P$ and $U_Q$ be unitary operators and $\rho_A$ a density matrix acting on $\mathcal{H}_A$. In this article, we implement a measurement-free version of a class of quantum operations defined by
	\begin{eqnarray}
		\label{mainO}
		O\left[ \rho_A\right]=
		U_Q Q\rho_A Q U_Q^\dag+U_P P \rho_A P U_P^\dag.
	\end{eqnarray}
	A standard measurement-based implementation of these operations would be to measure an observable corresponding to the projectors $P$ and $Q$, and depending on the outcome, implement a unitary $U_P$ or $U_Q$.
	
	Although these operations are defined for a general $N$-dimensional system, it includes a number of illustrative examples for a qubit $N=2$ with states $|0\rangle, |1\rangle$. Defining $P=|1\rangle \langle 1|$, $U_P=\sigma_x$ and $U_Q=1$, we implement
	\begin{eqnarray}
		\label{decayO}
		O_{\text{decay}}\left[ \rho_A\right]
		=\begin{pmatrix}
			1 & 0 \\ 0 & 0
		\end{pmatrix}.
	\end{eqnarray}
	On the other hand, a dephasing operation can be defined similarly with $P=|1\rangle \langle 1|$, $U_P=1$ and $U_Q=1$ so that 
	\begin{eqnarray}
		\label{dephasingO}
		O_{\text{dephasing}}\left[\rho_A \right]=\begin{pmatrix}
			(\rho_A)_{0,0} & 0 \\ 0 &  (\rho_A)_{1,1}
		\end{pmatrix}.
	\end{eqnarray}
	As an example of a phase preserving operation, when $P=1$ and $U_P=U$, we implement the general unitary
	\begin{eqnarray}
		\label{unitary}
		O_{\text{unitary}}\left[\rho_A \right]=U \rho_A U^\dag
		\label{UnitaryO}
	\end{eqnarray}
	as a steady state. 
	
	The class of operations in Eq. \eqref{mainO} cannot be implemented as a map from an initial state to a steady state of a time-independent Lindblad operator acting directly on the system. For a time-independent Lindblad operator $\mathcal{L}$, this would require $\mathcal{L}[O[\rho_A]]=0$ \textit{for all} $\rho_A$. We show below how this assumption leads to a contradiction. 
	
	Although here we implement phase-preserving operations, there is an important caveat: phase preservation requires finely tuned Hamiltonian couplings. We study the robustness of this phase preservation below and find that it survives moderate disorder.

	\section{Spectral Switching}
	We now implement these operations via coupling to an ancilla. We define the Hilbert space $\mathcal{H}=\mathcal{H}_A \otimes \mathcal{H}_B$, where $\mathcal{H}_A$ is the Hilbert space of the main system and $\mathcal{H}_B$ is for the ancillary states. To reach a steady state in $\mathcal{H}_A$, the ancilla $\mathcal{H}_B$ must be infinite in size, otherwise any initial state will return arbitrarily close to itself eventually, up to a global phase \cite{recurrencePoincare,recurrenceNote,recurrence2026,Recurr}. The full Hamiltonian is   $H=H_A+H_{\text{int}}$ where $H_A$ is the main subsystem's Hamiltonian, and $H_{\text{int}}$ is the main system-ancilla interaction.

	We start by implementing a base case of Eq. \eqref{mainO} where $P=|i\rangle \langle i |$ is a rank-1 projection and $U_Q=1$ acts trivially. Then our goal is to enact a unitary on just the state $|i\rangle$, mapping it to the state $|U_Pi\rangle=U_P|i\rangle$. We introduce the Hamiltonian $H=H_A+H_{\text{int}}$ that implements this. We first define the Hamiltonian of the main system as
	\begin{eqnarray}
		\label{HA}
		H_A=h_A \otimes 1_B
	\end{eqnarray}
	where $h_A$ acts in $\mathcal{H}_A$ and $1_B$ is the identity operator in $\mathcal{H}_B$. The ancilla $\mathcal{H}_B$ is infinite dimensional with orthonormal basis states $|j\rangle=|0\rangle, \ |1\rangle, \ |2\rangle, \ \dots$.  We choose a simple $h_A=E_P P+E_{Q} Q=E_P |i \rangle \langle i |+E_{Q} Q$. If we want to have $\langle i |U_Pi\rangle \neq 0$ (such as in the dephasing case), then we require $E_P=E_{Q}$. If we have orthogonality $\langle i |U_Pi\rangle=0$ (such as in the decay case), this is not a requirement of $H_A$.
	
	We define the main system-ancilla interaction as 
	\begin{eqnarray}
		\label{int}
		H_{\text{int}}=&C& |U_Pi \rangle \langle i | \otimes |1\rangle \langle 0 | +\text{H.c.} \nonumber \\ &+&B\sum_{j=1}^\infty ( |U_Pi \rangle \langle U_Pi | \otimes |j+1\rangle \langle j | +\text{H.c.})
	\end{eqnarray}
	where $C$ and $B$ are coupling constants. The idea is for the state to move in a chainlike way from  $|x_0\rangle=|i\rangle \otimes |0\rangle$ to $|x_1\rangle=|U_Pi\rangle \otimes |1\rangle$ to $|x_2\rangle=|U_Pi\rangle \otimes |2\rangle$ etc. While $H_{\text{int}}$ implements this process, the system Hamiltonian $H_A$ will allow or suppress this process through spectral resonance.
	
	Since $H$ acts nontrivially only on the $|x_j\rangle$ states, we define $H_X=1_X H 1_X$ for the projector $1_X=\sum_j |x_j\rangle \langle x_j|$. The $|x_j\rangle$-projected Hilbert space is $\mathcal{H}_X \subset \mathcal{H}$. This allows us to utilize a general matrix theory, where we represent $|x_j\rangle$ as standard orthonormal column vectors. We define the energy difference $\mu=E_P-E_Q$ and the normalized values $\mu_B=\mu/B$ and $C_B=C/B$. Then we have $H_X = E_Q 1_X + 2Bh_X$ where  
	\begin{eqnarray}
		\label{chainMapping}
		h_X=\begin{pmatrix}
			\mu_B/2 & C_B/2 & 0 & 0 & \dots \\
			C_B/2 & 0 & 1/2 & 0 & \dots \\
			0 & 1/2 & 0 & 1/2 & \dots \\
			0 & 0 & 1/2 & 0 & \dots \\
			\vdots & \vdots & \vdots & \vdots & \ddots \\
		\end{pmatrix}.
	\end{eqnarray}
	The Hamiltonian $h_X$ has an absolutely continuous spectrum in $[-1,1]$. Depending on the values $\mu_B$ and $C_B$, there may also be discrete energy levels outside of this range. Therefore, we present a spectral criterion for $h_X$ to \textit{not} have discrete energy levels. For $C_B \neq 1$, the spectral criterion is
	\begin{eqnarray}
		\label{decay}
		2\left|1-C_B^2\right|<\left| |\mu_B|-\sqrt{\mu_B^2+4C_B^2-4} \right|.
	\end{eqnarray}
	If $C_B=1$, the condition is $|\mu_B|<1$ and if $\mu_B=0$ the condition is $C_B<\sqrt{2}$. 
	
	The model $h_X$ maps to a special case of the Anderson impurity model \cite{Anderson}, where the discrete energy levels correspond to localized bound states on a lattice. Equivalently, $h_X$ is a Fano-Anderson (Friedrichs) model, in which a discrete level is coupled to a band of finite width. The criterion \eqref{decay} is the condition for no bound state to split off from that band. The same mechanism underlies the atom-photon bound states of waveguide and photonic-bandgap QED, where an emitter tuned inside the band decays while an emitter tuned outside it does not \cite{JohnWang1990,Calajo2016,Longhi2007}. What is new here is not this spectral dichotomy itself, but its use as a programmable selector. We show below that it switches on and off an entire class of measurement-and-feedforward operations under a single time-independent Hamiltonian.

	We  now show how these spectral properties relate to the dynamics of operators, in particular 
	$n_j=|x_j \rangle \langle x_j|$. For this operator, we state the following theorem, which provides a condition to guarantee a steady state. \\

	\noindent \textbf{Spectral switching theorem} \textit{Consider the Hamiltonian  $H_X = E_Q 1_X + 2Bh_X$ with parameters within the spectral criterion \eqref{decay}. Then for any initial state $|\psi_x \rangle \in \mathcal{H}_X$, the time-evolved expectation values $\langle{n}_j (t)\rangle$ have the limit
		\begin{eqnarray}
			\lim_{t \to +\infty} \langle{n}_j (t)\rangle = 0.
	\end{eqnarray}}
	\\ \noindent This theorem is the main mathematical tool used throughout this article to guarantee steady state operations of the form Eq. \eqref{mainO}. This is rigorously proven in Appendix \ref{appendixA} using spectral measure theory \cite{spectra, spectra2}. It is named the spectral switching theorem because the decay $\langle n_j(t) \rangle \to 0$ is guaranteed within the spectral criterion \eqref{decay}, whereas outside that criterion a bound state splits off from the band and $\langle n_0(t)\rangle$ retains a finite long-time value, computed in closed form in Appendix \ref{appendixB}. The operation is thus executed within the spectral criterion and suppressed outside of it. 
	
	\color{black}
	Using the spectral switching theorem, we first prove our implementation of Eq. \eqref{mainO}  for the case when the projector $P=|i\rangle \langle i|$ is rank 1 and $U_Q=1$ is trivial. We start from 
	\begin{eqnarray}
		|\psi(t=0)\rangle&=&|\psi_A\rangle \otimes |0\rangle 
		=(Q+P)|\psi_A\rangle \otimes |0\rangle \nonumber \\
		&=&Q|\psi_A\rangle \otimes |0\rangle+\langle i|\psi_A\rangle |x_0 \rangle
	\end{eqnarray}
	In the state above, the Hamiltonian only acts non-trivially on the $|x_0\rangle$ component. Thus we define $\chi_j(t)$ through $e^{-iH_X t} |x_0\rangle = \sum_j \chi_j(t) |x_j\rangle$. This implies that $|\chi_j(t)|^2=\langle n_j(t)\rangle$. We get

	\begin{eqnarray}
		|\psi(t)\rangle&=&e^{-iE_Q t}Q|\psi_A\rangle \otimes |0\rangle+\langle i|\psi_A\rangle \sum_j \chi_j(t) |x_j \rangle \nonumber \\
		&=&\left(e^{-iE_Q t} Q|\psi_A\rangle+\langle i|\psi_A\rangle \chi_0(t) |i\rangle\right) \otimes |0\rangle \nonumber \\ &+& \langle i|\psi_A\rangle\sum_{j>0} \chi_j(t) |U_Pi\rangle \otimes |j\rangle  
	\end{eqnarray} 
	Now we define $\rho(t)=|\psi(t)\rangle \langle \psi(t)|$ and the reduced density matrix on $\mathcal{H}_A$ as $\rho_A(t)=\Tr_B[\rho(t)]$. We get
	\begin{eqnarray}
		\rho_A(t)&=& \left(e^{-iE_Q t}Q|\psi_A\rangle+\langle i|\psi_A\rangle \chi_0(t) |i\rangle\right) \nonumber \\ &\times& \left(\langle \psi_A|Qe^{iE_Q t}+\langle \psi_A| i \rangle \chi_0^*(t) \langle i|\right)\nonumber \\&+&(1-\langle n_0(t) \rangle) |\langle i|\psi_A\rangle|^2 |U_Pi \rangle \langle U_Pi|. 
	\end{eqnarray} 
	where we have used  $\sum_{j>0}|\chi_j(t)|^2=1-\langle n_0(t)\rangle$. Due to the spectral switching theorem, $|\chi_0(t\to \infty)|^2=\langle n_0(t\to \infty)\rangle=0$. Thus
	\begin{eqnarray}
		\rho_A(t\to \infty)&=&Q|\psi_A\rangle \langle \psi_A|Q+|\langle i|\psi_A\rangle|^2|U_Pi\rangle \langle U_Pi| \nonumber \\
		&=& Q|\psi_A\rangle \langle \psi_A|Q+U_P P|\psi_A \rangle \langle \psi_A| P U_P^\dag. \nonumber \\
	\end{eqnarray}
	Using linearity, this generalizes to starting from the mixed state $\rho(t=0)=\rho_{A,0} \otimes |0\rangle \langle 0|$ to
	\begin{eqnarray}
		\rho_A(t\to \infty) &=& Q\rho_{A,0}Q+U_P P\rho_{A,0} P U_P^\dag. \nonumber \\
	\end{eqnarray}
	Thus we have implemented the operation Eq. \eqref{mainO} for the rank-1 projector $P=|i\rangle \langle i |$ and trivial $U_Q=1$, which occurs for the spectral criterion Eq. \eqref{decay}.

	\subsection*{Example}
	Although this work is a broad framework for spectral control, a useful example is a spin-qubit coupled to a one-dimensional ferromagnetic spin chain. In particular, we consider the one-dimensional XY model
	\begin{eqnarray}
		H=\mu \sigma_1^z + C (\sigma_1^+ \sigma_2^-+\text{H.c.}) 
		+B\sum_{j=2}^L (\sigma_j^+ \sigma_{j+1}^-+\text{H.c.}). \nonumber \\
	\end{eqnarray}
	Here $\mu$ represents an on-site magnetic field while $C$ and $B$ represent spin-spin interactions. The first spin is  the qubit while spins $j\geq2$ are ancilla spins. We start the ancilla in the spin down state $|0\rangle=|\downarrow \downarrow \downarrow \dots \downarrow \rangle$ while the first spin can be in any initial state. Mathematically, we start the system in the state $\rho(t=0)=\rho_{A,0} \otimes |0\rangle \langle 0|$ where $\rho_{A,0}$ is the 2$\times$2 density matrix of the first spin. 
	
	The goal, then, is to reset the qubit to the state $|\downarrow \rangle \langle \downarrow|$. We define the $|x_j\rangle$ states and the intended hopping process as $|x_0\rangle=|\uparrow \rangle \otimes |\downarrow \downarrow \downarrow \downarrow \dots \downarrow \rangle$ to $|x_1\rangle=|\downarrow \rangle \otimes |\uparrow \downarrow \downarrow \downarrow \dots \downarrow \rangle$ to $|x_2\rangle=|\downarrow \rangle \otimes |\downarrow \uparrow \downarrow \downarrow \dots \downarrow \rangle$ to $|x_3\rangle=|\downarrow \rangle \otimes |\downarrow \downarrow \uparrow \downarrow \dots \downarrow \rangle$ etc. The process thus describes an up-spin hopping through the ancilla as a magnonic excitation. The part of the state that starts in $|\downarrow \rangle \otimes | \downarrow \downarrow \downarrow \dots  \downarrow \rangle$ is an eigenstate of $H$ and only evolves through a dynamical phase. When $L \to \infty$, the action of $H$ on the $|x_j\rangle$ states maps directly on
	to the $H_X$ Hamiltonian of Eq. \eqref{chainMapping}.  Therefore the occupation of $|x_0 \rangle$ goes to $0$ in the long-time limit. Since all other $|x_j\rangle$ have down spins for the qubit, the qubit is therefore reset to the  $|\downarrow\rangle$ state. 
	
	\begin{figure}
		\centering
		\includegraphics[width=0.49\linewidth]{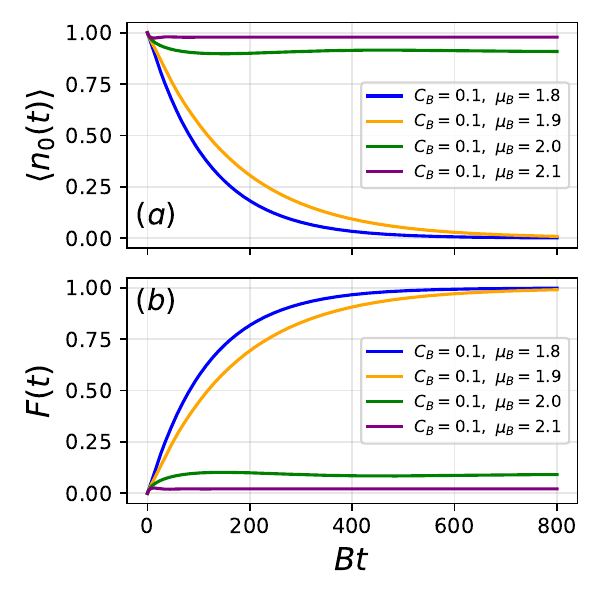} \includegraphics[width=0.49\linewidth]{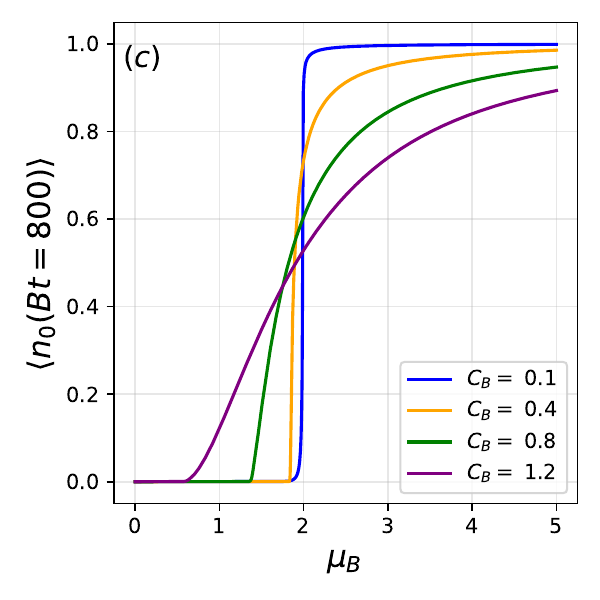}
		\caption{\color{black}Demonstration of the spectral switch through time evolution under Hamiltonian $H_X=E_Q 1_X+2 B h_X$ with $h_X$ in Eq. 
			\eqref{chainMapping}. (a) $\langle n_0 (t) \rangle$ and (b) fidelity $F(t)$ for $C_B=0.1$ and various $\mu_B$. The spectral criterion Eq. \eqref{decay} is met for $\mu_B \lesssim 1.99$. The state resets $F(t)=1$ when the spectral criterion is met. (c) The long time limit for various parameters. The switch is sharpest for small $C_B$.} 
		\label{fig:switch}
	\end{figure}
	
	We numerically analyze this example for finite ancilla sizes and demonstrate that the Hamiltonian reliably acts as a spectral switch over a finite time window. As an example, we start the system in  $\rho(t=0)=|\uparrow \rangle \langle\uparrow| \otimes |0\rangle \langle 0 |=|x_0\rangle \langle x_0|$. Figure~\ref{fig:switch}a shows the occupation of $|x_0\rangle$ given by $\langle n_0(t)\rangle$ for system size $L=900$, and parameters $B=1$, $C=0.1$, $E_Q=0$, and $\mu_B=1.8,1.9,2.0,2.1$. For these parameters the spectral criterion Eq.~\eqref{decay} gives $\mu_B < 2\sqrt{1-0.01}\approx 1.99$, so the $\mu_B=1.8$ and $1.9$ cases decay $\langle n_0\rangle$ to $0$ while the others do not. We use the fidelity of $\rho_A(t)$ with respect to the reset state $\rho_{\downarrow}=|\downarrow\rangle\langle \downarrow|$, given by
	\begin{eqnarray}
		\label{resetFidelity}
		F(t) = \left( \Tr[\sqrt{\sqrt{\rho_A(t)} \rho_{\downarrow} \sqrt{\rho_A(t)}}] \right)^2.
	\end{eqnarray}
	$F(t)=1$ indicates that the state has decayed to the pure state $\rho_{\downarrow}$. The fidelity is plotted in Figure~\ref{fig:switch}b. Figure~\ref{fig:switch}c shows the long-time limit of $\langle n_0\rangle$ as a function of $\mu_B$ for several values of $C_B$, indicating that the spectral switch is sharpest for small $C_B$. The long-time limit of $\langle n_0(t)\rangle$ plotted in Figure~\ref{fig:switch}c is derived analytically in Appendix \ref{appendixB} and reads 
	\begin{eqnarray}
		\label{closedFormMain}
		&&\langle n_0(t\rightarrow \infty)\rangle= \Big( 1- \nonumber\\ && \frac{C_B^2(|\mu_B|-\sqrt{4C_B^2+\mu_B^2-4})^2}{(1-C_B^2)(4(1-C_B^2)-(|\mu_B|-\sqrt{4C_B^2+\mu_B^2-4})^2)}  \Big)^2. \nonumber \\
	\end{eqnarray} 
	As $C_B \rightarrow 0$ the spectral switch becomes sharpest: in this limit $\langle n_0(t\rightarrow \infty)\rangle=0$ for $|\mu_B|<2$, while $\langle n_0(t\rightarrow \infty)\rangle=1$ for $|\mu_B|>2$. This closed-form result quantifies analytically the sharpening seen numerically in Figure~\ref{fig:switch}c.
	
\begin{figure}
		\centering    \includegraphics[width=0.95\linewidth]{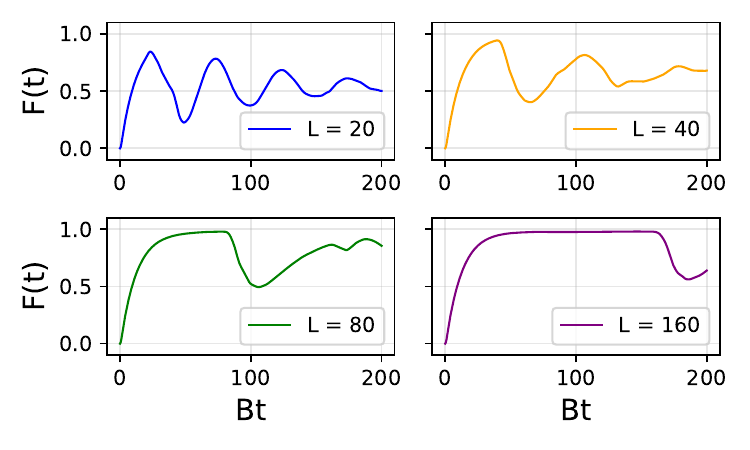}
		\caption{Emergence of a steady state with increased ancilla system size. Plotted is the fidelity in Eq. \eqref{resetFidelity} between the reset state and the time-evolved initial state. Parameters are $\mu_B=0.3$, $C_B=0.2$ and disorder $d_j \in [-0.1,0.1]$; we average 100 disorder realizations. The state is reset when $F(t)=1$.}
		\label{fig:resetFidelity}
	\end{figure}

	The above plots are for an ancilla system size $L$ large enough to prevent recurrence within the shown time window. We illustrate the system size dependence for  ancilla sizes $L=20,40,80,160$. We also include disorder 
	\begin{eqnarray}
		H_{\text{disorder}}= \sum_{j=0}^{L-1} d_j \left( 1_q \otimes |j \rangle \langle j | \right) 
	\end{eqnarray}
	where $1_q$ is  identity in the qubit space, $d_j$ is chosen from a uniform distribution in $[-0.1,0.1]$ and $\rho_A(t)$ is averaged over $100$ disorder realizations. We simulate a parameter set within the spectral criterion $B = 1$, $C = 0.2$, $E_P=0.3$ and $E_Q =0.0$.
	Figure \ref{fig:resetFidelity} shows the corresponding $F(t)$, demonstrating how the steady state emerges with increased ancilla system size. This numerical demonstration shows that an autonomous reset can be implemented for a finite window of time in a finite system.

	\section{Phase preserving steady states}
	\subsection{Generalizing the operation}
	\label{sec:general}
	Now we generalize the implementation of Eq. \eqref{mainO}. The simplest implementation is to remove the spectral control so that $H_A=0$ and only use a system-ancilla interaction. We extend the ancilla Hilbert space $\mathcal{H}_B$ to have positive and negative indices $|j\rangle=|0\rangle, |\pm1\rangle,|\pm2\rangle, \dots$ Then we use
	\begin{eqnarray}
		\label{intGeneral}    H_{\text{int}}&=&C_P \sum_{i\in S_P} (|U_Pi \rangle \langle i | \otimes |1\rangle \langle 0 | +\text{H.c.}) \nonumber \\ &+&B_P \sum_{i\in S_P} \sum_{j=1}^\infty ( |U_Pi \rangle \langle U_Pi | \otimes |j+1\rangle \langle j |  +\text{H.c.}) \nonumber \\
		&+&C_Q \sum_{i\notin S_P} (|U_Q i \rangle \langle i |  \otimes |-1\rangle \langle 0 |+\text{H.c.}) \nonumber \\ &+&B_Q \sum_{i\notin S_P} \sum_{j=1}^\infty ( |U_Qi \rangle \langle U_Qi |  \otimes |-j-1\rangle \langle -j |+\text{H.c.}). \nonumber \\
	\end{eqnarray} 
	For every  $|\psi_A\rangle \in \mathcal{H}_A$, define $|\psi_A^P\rangle=\frac{1}{\sqrt{\langle \psi_A|P|\psi_A \rangle}} P |\psi_A\rangle$ and $|\psi_A^Q\rangle=\frac{1}{\sqrt{\langle \psi_A|Q|\psi_A \rangle}} Q |\psi_A\rangle$. Then let
	\begin{eqnarray}
		|x_0^{\psi_A}\rangle= |\psi_A^P\rangle \otimes |0\rangle  \ &,& \  |x_j^{\psi_A}\rangle= U_P|\psi_A^P\rangle \otimes |j\rangle  \nonumber \\ |y_0^{\psi_A}\rangle= |\psi_A^Q\rangle \otimes |0\rangle  \ &,& \  |y_j^{\psi_A}\rangle= U_Q|\psi_A^Q\rangle \otimes  |-j\rangle  \nonumber \\
	\end{eqnarray} 
	for $j>0$. For a given $|\psi_A\rangle$, the states $\{|x_j^{\psi_A}\rangle\}$ and $\{|y_j^{\psi_A}\rangle\}$ span two invariant subspaces of $H_{\text{int}}$. Defining the corresponding projectors $1_X=\sum_j |x_j^{\psi_A}\rangle \langle x_j^{\psi_A}|$ and $1_Y=\sum_j |y_j^{\psi_A}\rangle \langle y_j^{\psi_A}|$, the restriction of $H_{\text{int}}$ to these subspaces decomposes as $H_X+H_Y$ with $H_X=1_X H_{\text{int}} 1_X$ and $H_Y=1_Y H_{\text{int}} 1_Y$. The action of $H_X$ (and similarly $H_Y$) on $|x_0^{\psi_A}\rangle$ ($|y_0^{\psi_A}\rangle$) is exactly described by the matrix \eqref{chainMapping} with couplings $C = C_P$, $B = B_P$ ($C=C_Q$, $B=B_Q$) and $\mu_B=0$. Then defining $n_0^X=|x_0^{\psi_A}\rangle \langle x_0^{\psi_A}|$ and $n_0^Y=|y_0^{\psi_A}\rangle \langle y_0^{\psi_A}|$, we can utilize the spectral switching theorem.

	We start in initial states
	\begin{eqnarray}
		|\psi(t=0)\rangle&=&|\psi_A\rangle \otimes |0 \rangle = (P+Q)|\psi_A\rangle \otimes |0\rangle \nonumber \\
		&=&p_0 |x_0^{\psi_A}\rangle+q_0 |y_0^{\psi_A}\rangle.
	\end{eqnarray}
	for constants $p_0$ and $q_0$ with $|p_0|^2=\langle \psi_A| P | \psi_A \rangle$ and $|q_0|^2=\langle \psi_A| Q | \psi_A \rangle$. Defining $e^{-iH_X t} |x_0\rangle = \sum_j \chi_j(t) |x_j\rangle$ and $e^{-iH_Y t} |y_0\rangle = \sum_j \upsilon_j(t) |y_j\rangle$, the time evolved state is given by
	\begin{eqnarray}
		|\psi(t)\rangle&=&p_0e^{-\im H_Xt}|x_0^{\psi_A}\rangle+q_0e^{-\im H_Y t}|y_0^{\psi_A}\rangle \nonumber \\
		&=&p_0\sum_j \chi_j(t)|x_j^{\psi_A}\rangle+q_0\sum_j \upsilon_j(t)|y_j^{\psi_A}\rangle.
	\end{eqnarray}
	We then define the full density matrix $\rho(t)=|\psi(t)\rangle \langle \psi(t)|$ and the reduced density matrix  $\rho_A(t)=\Tr_B[\rho(t)]$, arriving at
	\begin{eqnarray}
		\rho_A(t)&=&(\chi_0(t) P+\upsilon_0(t) Q)|\psi_A\rangle \langle \psi_A | (P\chi_0^*(t)+ Q\upsilon_0^*(t)) \nonumber \\ &+&(1-|\chi_0(t)|^2)U_P P |\psi_A\rangle \langle \psi_A | P U_P^\dag \nonumber  \\ &+&(1-|\upsilon_0(t)|^2)U_Q Q |\psi_A\rangle \langle \psi_A|  Q U_Q^\dag. 
	\end{eqnarray} 
	When the spectral criterion applies, $|\chi_0(t \rightarrow \infty)|^2 =\langle n_0^X(t \rightarrow \infty) \rangle= 0$ and $|\upsilon_0(t \rightarrow \infty)|^2 =\langle n_0^Y(t \rightarrow \infty) \rangle= 0$, and thus 
	\begin{eqnarray}
		\rho_A(t\rightarrow \infty)=U_P P |\psi_A\rangle \langle \psi_A | P U_P^\dag +U_Q Q |\psi_A\rangle \langle \psi_A|  Q U_Q^\dag. \nonumber 
	\end{eqnarray} 
	Using linearity, this generalizes similarly to starting from the mixed state $\rho(t=0)=\rho_{A,0} \otimes |0\rangle \langle 0|$.

	\subsection{Non-Lindbladian proof}
	We show now that the general operation Eq. \eqref{mainO} cannot be implemented as a map from an initial state to a steady state of a time-independent Lindblad operator. To prove this, we need only to show that a subset of operations of $O[\rho_A]$ cannot be implemented; hence we consider a subset of operations. The argument below demonstrates a limitation of time-independent Markovian dissipation as an initial-to-steady-state map. This is exactly the gap that the Hamiltonian construction of Sec.~\ref{sec:general} fills constructively.
	
	We can form a subset of operations by placing restrictions on $U_P$ and $U_Q$ in Eq. \eqref{mainO}. We choose a case where $U_P$ and $U_Q$ are block diagonal in the $P$ and $Q$ subspaces of $\mathcal{H}_A$. Explicitly, we write the operators as
	\begin{eqnarray}
		\label{conditions}
		P=1_P \oplus 0_Q \ , \ Q=0_P \oplus 1_Q \nonumber \\
		U_P=u_P \oplus 1_Q \ , \ U_Q=1_P \oplus u_Q
	\end{eqnarray}
	where $1_P$ and $1_Q$ are identity operators, $0_P$ and $0_Q$ are zero operators and $u_P$ and $u_Q$ are unitary operators in the $P$ and $Q$ subspaces, respectively. We  now assume that $O[\rho_A]$ in Eq. \eqref{mainO} with conditions in \eqref{conditions} is implementable as a map from an initial state to a steady state of a Lindblad operator. Next we prove this requires a trivial case with $u_P=1_P$ and $u_Q=1_Q$; therefore selecting any $u_P \neq 1_P$ or $u_Q\neq1_Q$ is outside of the Lindblad class. 
	
	A Lindblad time evolution is given by $\rho_A(t)=\exp(\mathcal{L}t)[\rho_A(0)]$ where $\mathcal{L}[ \ \cdot \ ]$ is linear and such that $\exp(\mathcal{L}t)$ is a completely positive, trace-preserving map for all time $t$. For a steady state $\rho_A^{ss}$, we require $\exp(\mathcal{L}t)[\rho_A^{ss}]=\rho_A^{ss}$. That is, steady states are fixed points which do not evolve under the time evolution operator. This implies $\mathcal{L}[\rho_A^{ss}]=0$; i.e. $\rho_A^{ss}$ is in the kernel of $\mathcal{L}$. Then to implement Eq. \eqref{mainO} as a steady state, we require
	\begin{eqnarray}
		\mathcal{L}\Big[ O[\rho_A] \Big]=0 
	\end{eqnarray}
	for all possible $\rho_{A}$. This includes $\rho_A=(u_P^\dag \rho_A^P u_P) \oplus 0_Q$ for all possible density matrices $\rho_A^P$ in the $P$ subspace. Plugging this into Eq. \eqref{mainO} with conditions \eqref{conditions}, we get
	\begin{eqnarray}
		\mathcal{L}[\rho_A^P \oplus 0_Q]=0.
	\end{eqnarray}
	for all possible $\rho_A^P$. Equivalently,
	\begin{eqnarray}
		\label{LindbladExp}
		\exp(\mathcal{L}t)[\rho_A^P \oplus 0_Q]=\rho_A^P \oplus 0_Q.
	\end{eqnarray}
	for all possible density matrices $\rho_A^P$ and for all time $t$. This means the Lindblad operation acts trivially on all density matrices of this form. On the other hand
	\begin{eqnarray}
		\label{ONL}
		O[\rho_A^P\oplus 0_Q]=(u_P \rho_A^P u_P^\dag)\oplus 0_Q.
	\end{eqnarray}
	Thus for Eq. \eqref{LindbladExp} to reach Eq. \eqref{ONL} in the long time limit, we require $u_P \rho_A^P u_P^\dag=\rho_A^P$ for every $\rho_A^P$, i.e. $u_P=1_P$ up to an irrelevant global phase. A similar proof shows that $u_Q=1_Q$ is also required. Thus so long as $u_P \neq 1_P$ or $u_Q \neq 1_Q$, the operation Eq. \eqref{mainO} with conditions \eqref{conditions} is outside of the class of Lindblad steady-state operations. 
	
	Here we stress the precise scope of this argument. It excludes time-independent Lindblad generators for which $O[\rho_A]$ is a fixed point for every $\rho_A$. It does not exclude time-dependent generators, nor a finite-time Markovian channel $\exp(\mathcal{L}t^*)$ evaluated at some finite $t^*$, both of which require external timing control of exactly the kind the present construction avoids. Furthermore, by expanding the Hilbert space to include a dissipative ancilla, operations of the form in Eq. \eqref{mainO} can be engineered via Lindblad jump operators $L_P=U_P P \otimes |0\rangle \langle 1|$ and $L_Q=U_Q Q \otimes |0\rangle \langle 1|$ acting on the expanded space \cite{AlbertJiang2}. However, as noted in the Introduction, this explicitly trades closed-system dynamics for reliance on the control of an external bath.
	
	\begin{figure}
		\centering    \includegraphics[width=0.9\linewidth]{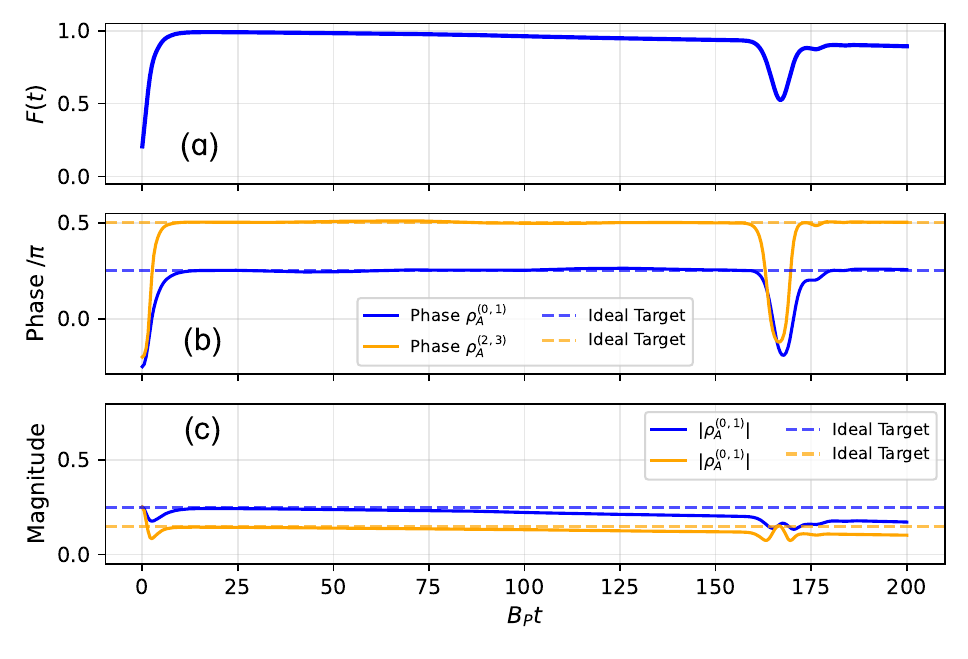}
		\caption{Phase-preserving steady state operation in the presence of disorder Eq. \eqref{phaseDisorder} with parameters in Sec. \ref{RobustPhase}. Plotted is the fidelity in Eq. \eqref{OFidelity} between the operation $O[\rho_A]$ and the time evolved initial state. (a) Here we plot the fidelity which slowly decreases due to the disorder. (b) The phase is largely preserved even in the presence of disorder. (c) The magnitude is decaying slowly as a function of time. At $B_Pt \approx 175$ the excitation has reached the boundary of the finite ancilla, ending the steady-state behavior. }
		\label{fig:Phase}
	\end{figure}
	
	\subsection{Robustness of phase preservation}
	\label{RobustPhase}
	Now we study the robustness of operations with the conditions \eqref{conditions} in this time-independent Hamiltonian formalism defined in Eq. \eqref{intGeneral}. We choose an example where $\mathcal{H}_A$ is a four-dimensional Hilbert space with states $|0\rangle, |1\rangle, |2\rangle, |3\rangle$. We choose operators $P=|0\rangle \langle 0 |+ |1\rangle \langle 1|=1_P \oplus 0_Q$, $Q=|2\rangle \langle 2 |+ |3\rangle \langle 3| = 0_P \oplus 1_Q$, $U_P=\sigma_x \oplus 1_Q$ and $U_Q=1_P \oplus (\frac{1}{\sqrt{2}}\sigma_x+\frac{1}{\sqrt{2}}\sigma_z)$. For parameters we set $C_P=0.5$, $C_Q=0.4$, $B_P=1$ and $B_Q=1$.
	
	To study the robustness of this operation, we look at a finite ancilla size of $L=160$ and include disorder by  adding a disorder Hamiltonian of the form
	\begin{eqnarray}
		\label{phaseDisorder}
		H_{\text{dis}}=\sum_{i,j} d_{i,j} |i\rangle \langle i| \otimes |j\rangle \langle j|
	\end{eqnarray}
	where each $d_{i,j}$ is drawn uniformly from $[-0.1,0.1]$ and we average $\rho_A(t)$ over 100 disorder realizations.
	
	We choose an initial pure state $|\psi(t=0)\rangle=|\psi_A\rangle \otimes |0\rangle$ with 
	\begin{eqnarray}
		|\psi_A\rangle=\frac{1}{2}(|0\rangle+e^{\im \pi/4}|1\rangle+e^{3\im \pi/5}|2\rangle+\im e^{\im 4\pi/5}|3\rangle). \nonumber
	\end{eqnarray}
	Defining $\rho_{A,0}=|\psi_A\rangle \langle \psi_A|$, we have the intended operation:
	\begin{eqnarray}
		O[\rho_{A,0}]= \frac{1}{4} \begin{pmatrix} 1 & e^{-\im \frac{\pi}{4}} \\ e^{\im \frac{\pi}{4}} & 1 \end{pmatrix} \oplus \begin{pmatrix} 1+ \cos\frac{\pi}{5} & -\im \sin\frac{\pi}{5} \\ \im \sin\frac{\pi}{5} & 1-\cos\frac{\pi}{5} \end{pmatrix}. \nonumber
	\end{eqnarray} 
	This operation thus preserves the relative phase information between $|0\rangle$ and $|1\rangle$ and between $|2\rangle$ and $|3\rangle$.
	
	We look at the fidelity of the time evolution 
	\begin{eqnarray}
		\label{OFidelity}
		F(t) = \left( \Tr[\sqrt{\sqrt{\rho_A(t)} O[\rho_{A,0}] \sqrt{\rho_A(t)}}] \right)^2
	\end{eqnarray}
	to see the overall quality of the operation. We also look at the specific phase terms $\rho_A^{(0,1)}=\langle 0 |\rho_A(t) |1\rangle$ and $\rho_A^{(2,3)}=\langle 2 |\rho_A(t) |3\rangle$. These parameters are plotted in Figure \ref{fig:Phase}. The disorder degrades the overall fidelity of the operation, and the magnitudes of the coherences decay slowly with time. The relative phases of $\rho_A^{(0,1)}$ and $\rho_A^{(2,3)}$, however, remain close to their target values throughout the window in which the steady state holds.  
	
	\section{Conclusion}
	In this work we have introduced a minimal, time-independent Hamiltonian that implements a class of measurement-and-feedforward operations as a map from an initial state to a local steady state of the system. Three features distinguish this construction. First, no measurement, classical feedback, or externally timed shut-off is required: a single static Hamiltonian, switched on once, carries out the operation. Second, the operation is spectrally addressable: whether it executes at all is controlled by the criterion \eqref{decay}, whose sharpness we obtain in closed form. Third, the steady states reached in this way include phase-preserving operations that provably cannot arise as initial-to-steady-state maps of any time-independent Lindblad equation acting \textit{only} on the system.
	
	There are two important costs to note clearly. First, the ancilla must be prepared in a definite initial state, so the protocol consumes a prepared resource in the same way that a cold bath does. Second, in a finite ancilla of size $L$ the steady state is only an intermediate-time plateau, valid until the excitation reaches the boundary, as seen in Figures~\ref{fig:resetFidelity} and~\ref{fig:Phase}. 
	
	Natural extensions include projectors of higher rank with several distinct target unitaries, ancillas with engineered densities of states that sharpen the switch further, and the question of whether error-correcting operations of practical interest fall within the class \eqref{mainO}. More broadly, it would be interesting to map out which nonunitary operations admit an autonomous, time-independent Hamiltonian implementation, and which genuinely require either measurement or external timing.
	\section*{Acknowledgments}This project was supported by the Max Planck--UBC--UTokyo Center for Quantum Materials, and the Natural Sciences and Engineering Research Council of Canada. 
	
	\section*{Data availability}
	The data that support the findings of this article are not publicly available. The data are available from the authors upon reasonable request.
	
	\appendix
	\section{Spectral Theory}
	\label{appendixA}
	The purpose of this appendix is to develop the spectral theory \cite{spectra, spectra2} needed for the spectral switching theorem.
	
	\subsection{Free Jacobi operator $\Delta$}
	The free Jacobi operator is a self-adjoint operator on the $\ell^2$ inner product space acting on the simple column vectors $|x_k\rangle$. These vectors are orthonormal in $\ell^2$, written as $\langle x_k | x_{k'} \rangle = \langle x_k, x_{k'} \rangle=\delta_{k,k'}$. The free Jacobi operator is defined as
	\begin{eqnarray}
		\Delta=\begin{pmatrix}
			0 & 1/2 & 0 & 0 & \dots \\
			1/2 & 0 & 1/2 & 0 & \dots \\
			0 & 1/2 & 0 & 1/2 & \dots \\
			0 & 0 & 1/2 & 0 & \dots \\
			\vdots & \vdots & \vdots & \vdots & \ddots \\
		\end{pmatrix}.
	\end{eqnarray}
	By definition, the spectrum of $\Delta$ is the set of real numbers $E$ such that $E\textbf{1}-\Delta$ is not invertible (where \textbf{1} is the identity). The spectrum of $\Delta$ is absolutely continuous in the range $[-1,1]$. 
	
	Spectral theory is particularly useful to analyze semi-infinite matrices like $\Delta$ because normalization is not straightforward for nonlocalized states in an infinite Hilbert space. The spectral measure of $\Delta$ is
	\begin{eqnarray}
		d \mu_\Delta(s)=\frac{2}{\pi} \sqrt{1-s^2} ds
	\end{eqnarray}
	on $s \in [-1,1]$ and $0$ otherwise. The function $\frac{2}{\pi}\sqrt{1-s^2}$, physically, is the density of states in energy. 
	
	Using this spectral measure, we define a new inner product on a Hilbert space $L_\mu^2$. The orthonormal polynomials $Q_k(s) \in L_{\mu}^2$ of $\Delta$ are the Chebyshev polynomials of the second kind. They are orthonormal in the sense that
	\begin{eqnarray}
		\langle Q_k(s), Q_{k'}(s) \rangle_{L_\mu^2}=\int_\mathbf{s \in R} Q_k^*(s) Q_{k'}(s) d \mu_{\Delta}(s)=\delta_{k,k'}. \nonumber \\
	\end{eqnarray}
	Then we define the unitary operator $U_\Delta: \ell^2 \rightarrow L_\mu^2$ through the relation $U x_k=Q_k(s)$. 
	
	The quantities $U_\Delta$ and $\Delta$ are related in the following sense: First, the product of operators $U_\Delta \Delta U_\Delta^\dag$ is a self-adjoint operator on $L_\mu^2$. Let $f(s)$ be any function in $L_\mu^2$. Then the quantities here are related by the spectral relation for $\Delta$, which is
	\begin{eqnarray}
		\label{UDeltafs}
		U_\Delta \Delta U_\Delta^\dag f(s)=s f(s).
	\end{eqnarray}
	\subsection{Finite rank perturbation $h$}
	Next,  we extend this theory to perturbations of the free Jacobi operator $\Delta$. Similar to $\Delta$, we define 
	\begin{eqnarray}
		h=\begin{pmatrix}
			\mu/2B & C/2B & 0 & 0 & \dots \\
			C/2B & 0 & 1/2 & 0 & \dots \\
			0 & 1/2 & 0 & 1/2 & \dots \\
			0 & 0 & 1/2 & 0 & \dots \\
			\vdots & \vdots & \vdots & \vdots & \ddots \\
		\end{pmatrix}.
	\end{eqnarray}
	Now the goal is to determine the spectrum, the spectral measure $d \mu$, and orthonormal polynomials $P_k(s)$ of $h$.
	
	To determine the spectral theory of $h$, we relate it to the spectral theory of $\Delta$ using a connection coefficient matrix formalism \cite{spectra}. We apply Corollary 3.3 of Ref. \cite{spectra} to $h$ and $\Delta$ and determine the connection coefficient \scriptsize
	\begin{eqnarray}
		\label{connectionCoef}
		\mathcal{C}=\begin{pmatrix}
			1 & -\mu/C & B/C-C/B & 0 & 0 & \dots \\
			0 & B/C & -\mu/C & B/C-C/B & 0 & \dots \\
			0 & 0 & B/C & -\mu/C & B/C-C/B  & \dots \\
			0 & 0 & 0 & B/C & -\mu/C &  \dots \\
			0 & 0 & 0 & 0 & B/C & \dots \\
			\vdots & \vdots & \vdots & \vdots & \vdots & \ddots \\
		\end{pmatrix}. \nonumber \\
	\end{eqnarray} \normalsize
	$\mathcal{C}$ has a Toeplitz $\mathcal{C}_{\text{Toe}}$ and finite rank component $\mathcal{C}_{\text{fin}}$. They are \scriptsize
	\begin{eqnarray}
		\mathcal{C}_{\text{Toe}}=\begin{pmatrix}
			B/C & -\mu/C & B/C-C/B & 0 & 0 & \dots \\
			0 & B/C & -\mu/C & B/C-C/B & 0 & \dots \\
			0 & 0 & B/C & -\mu/C & B/C-C/B  & \dots \\
			0 & 0 & 0 & B/C & -\mu/C &  \dots \\
			0 & 0 & 0 & 0 & B/C & \dots \\
			\vdots & \vdots & \vdots & \vdots & \vdots & \ddots \\
		\end{pmatrix} \nonumber \\ 
	\end{eqnarray} \normalsize
	and
	\begin{eqnarray}
		\mathcal{C}_{\text{fin}}&=&\begin{pmatrix}
			(1-B/C) & 0 & 0 & 0 & 0 & \dots \\
			0 & 0 & 0 & 0 & 0 & \dots \\
			0 & 0 & 0 & 0 & 0  & \dots \\
			0 & 0 & 0 & 0 & 0 &  \dots \\
			0 & 0 & 0 & 0 & 0 & \dots \\
			\vdots & \vdots & \vdots & \vdots & \vdots & \ddots \\
		\end{pmatrix}. \nonumber \\
	\end{eqnarray} 
	so that $\mathcal{C}=\mathcal{C}_{\text{Toe}}+\mathcal{C}_{\text{fin}}$. The `Toeplitz symbol' of $\mathcal{C}_{\text{fin}}$ is 
	\begin{eqnarray}
		\bar{\mathcal{C}}(e^{\im \theta})=\frac{B}{C}-\frac{\mu}{C} e^{\im \theta}+(\frac{B}{C}-\frac{C}{B})e^{2 \im \theta}.
	\end{eqnarray}
	With $s=\cos \theta$, the magnitude $|\bar{\mathcal{C}}(e^{\im \theta})|^2$ can be expressed in terms of the second-order Chebyshev polynomials $Q_0(s)=1$, $Q_1(s)=2 \cos \theta$ and $Q_2(s)=4 \cos^2 \theta -1$. That is, 
	\begin{eqnarray}
		\label{Csquared}
		|\bar{\mathcal{C}}(e^{\im \theta})|^2=\sum_{k=0}^2\langle \mathcal{C}^T e_k, \mathcal{C}^T e_0 \rangle Q_k(s).
	\end{eqnarray}
	where the inner products are 
	\begin{eqnarray}
		\label{usefulCoefInner}
		\langle \mathcal{C}^T e_0, \mathcal{C}^T e_0 \rangle&=&1+(\frac{\mu}{C})^2+(\frac{B}{C}-\frac{C}{B})^2 \nonumber \\
		\langle \mathcal{C}^T e_1, \mathcal{C}^T e_0 \rangle&=&-\frac{\mu}{C}(\frac{2B}{C}-\frac{C}{B}) \nonumber \\
		\langle \mathcal{C}^T e_2, \mathcal{C}^T e_0 \rangle &=&\frac{B}{C}(\frac{B}{C}-\frac{C}{B}).
	\end{eqnarray}  
	Using the connection coefficient $\mathcal{C}$ of Eq. \eqref{connectionCoef}, we can determine the spectral theory of $h$. In particular, we only consider the case with an absolutely continuous spectrum on $[-1,1]$, with no discrete spectrum. The condition for this is the following: $|\bar{\mathcal{C}}(Z)|^2$ has no zeros on the unit disk $|Z| < 1$. This condition is equivalent to
	\begin{eqnarray}
		\label{decayConditionAppendix}
		2\left|1-C_B^2\right|<\left| |\mu_B|-\sqrt{\mu_B^2+4C_B^2-4} \right|
	\end{eqnarray}
	where $\mu_B=\mu/B$ and $C_B=C/B$. If $C=B$, the condition is $|\mu|<1$. Then the spectral measure of $h$ is 
	\begin{eqnarray}
		d \mu = \frac{d \mu_\Delta}{|\bar{\mathcal{C}}(e^{\im \theta})|^2}.
	\end{eqnarray}
	The orthonormal polynomials of $h$ are
	\begin{eqnarray}
		\label{Psum}
		P_k(s)=\sum_{j=0}^k \mathcal{C}_{j,k} Q_j(s)
	\end{eqnarray}
	where $Q_k(s)$ are the orthonormal polynomials of the free Jacobi operator $\Delta$. For $f(s) \in L_\mu^2$ and $P_k(s)=U_h |x_m\rangle$, we have the spectral relation
	\begin{eqnarray}
		U_h h U_h^\dag f(s)=s f(s).
	\end{eqnarray}

	\subsection{Time evolution of expectation values}
	Here we look at the time evolution under $h$. $h$ maps exactly to the Hamiltonian $h_X$ of the main article, relevant for the spectral switching theorem. 
	
	The time evolution of an operator $\hat{A}$ can be written in terms of sums of inner products of the form
	\begin{eqnarray}
		\langle \hat{A}(t) \rangle_{n,m} =\langle e^{\im h t} x_n, \hat{A} e^{\im h t} x_m \rangle_{\ell^2}.
	\end{eqnarray}
	To calculate these type of operators, we have to transform the operators and states into the $L_\mu^2$ inner product space. First, we have by definition 
	\begin{eqnarray}
		U_h x_m = P_k(s)
	\end{eqnarray}
	where $P_k(s)$ is given by sums of Chebyshev polynomials of the second kind as in Eq. \eqref{Psum}. Then for the time evolution operator, we have
	\begin{eqnarray}
		U_h e^{-\im ht} U_h^\dag =e^{-\im s t}.
	\end{eqnarray}
	For operators, we consider the on-site occupation $n_j=|x_j \rangle \langle x_j|$. 
	
	We start with looking at $n_j$. Then with $\theta$ defined through $\cos \theta=s$, we have
	\begin{eqnarray}
		\label{njInner}
		\langle n_j(t) \rangle_{n,m} &=&\langle e^{-\im h t} x_n, n_j e^{-\im h t} x_m \rangle_{\ell^2} \nonumber \\ &=&\langle e^{-\im h t} x_n, x_j \rangle_{\ell^2} \langle x_j, e^{-\im h t} x_m \rangle_{\ell^2} 
	\end{eqnarray}
	Then using Eq. \eqref{Csquared}, we can calculate these inner products as
	\begin{eqnarray}
		\label{njIntegrals}
		\langle e^{-\im h t} x_n, x_j \rangle_{\ell^2} =\frac{2}{\pi} \int_{-1}^1 \frac{e^{\im s t} P_n^*(s) P_j(s) \sqrt{1-s^2}}{\sum_{k=0}^2\langle \mathcal{C}^T x_k, \mathcal{C}^T x_0 \rangle Q_k(s)} ds \nonumber \\
		\langle x_j, e^{-\im h t} x_m \rangle_{\ell^2}=\frac{2}{\pi}\int_{-1}^1 \frac{e^{-\im s t} P_j^*(s) P_m(s) \sqrt{1-s^2}}{\sum_{k=0}^2\langle \mathcal{C}^T x_k, \mathcal{C}^T x_0 \rangle Q_k(s)} ds \nonumber \\
	\end{eqnarray}

	\subsection{Proof of spectral switching theorem}
	Here we show that $\langle n_j(t) \rangle_{0,0}$ of Eq. \eqref{njInner} goes to $0$ in the limit $t \rightarrow \infty$. We prove an even stronger statement that the inner product $\lim_{t\rightarrow \infty}\langle e^{-\im h t} x_n, x_j \rangle_{\ell^2}=0$ (and similarly for $\langle x_n, e^{-\im h t} x_j \rangle_{\ell^2}$). Using the form of the integral of Eq. \eqref{njIntegrals}, we define
	\begin{eqnarray}
		I(s)= \frac{2 e^{\im s t} P_n^*(s) P_j(s) \sqrt{1-s^2}}{\pi \sum_{k=0}^2\langle \mathcal{C}^T x_k, \mathcal{C}^T x_0 \rangle Q_k(s)}
	\end{eqnarray}
	for $s \in [-1,1]$ and $I(s)=0$ otherwise. 
	
	Then since $|\bar{\mathcal{C}}(e^{\im \theta})|^2 =\sum_{k=0}^2\langle \mathcal{C}^T x_k, \mathcal{C}^T x_0 \rangle Q_k(s) \neq 0$, we know that $I(s)$ is a bounded function. Furthermore it also only has support on $[-1,1]$. Therefore, we have that 
	\begin{eqnarray}
		\int_{-\infty}^\infty ds |I(s)|<\infty.
	\end{eqnarray}
	Then by the Riemann–Lebesgue lemma, we have $\lim_{t\rightarrow \infty}\langle e^{-\im h t} x_n, x_j \rangle_{\ell^2}=0$.
	
	\section{Bound-state properties}
	\label{appendixB}
	In this section, we consider a simpler analysis to study the properties of the bound state. We derive an expression for the long-term expectation value outside of the spectral criterion. We start with the eigenvalue equation
	\begin{eqnarray}
		\frac{1}{2}\begin{pmatrix} \mu_B & C_B & 0 & 0 & 0 & \dots \\ C_B & 0 & 1 & 0 & 0 &\dots \\ 0 & 1 & 0 & 1 & 0 &\dots \\ 
			0 & 0 & 1 & 0 & 1 &\dots \\
			\vdots & \vdots & \vdots & \vdots & \vdots  &\ddots \\
		\end{pmatrix} \begin{pmatrix} v_1 \\ v_2 \\ v_3 \\ v_4 \\ \vdots \\ \end{pmatrix}=\frac{E}{2}\begin{pmatrix} v_1 \\ v_2 \\ v_3 \\ v_4 \\ \vdots \\ \end{pmatrix}
	\end{eqnarray}
	
	Assume $v_2=A$, $v_3={Ar}$, $v_4=Ar^2$, etc. The bulk requirements of the solution are
	\begin{eqnarray}
		E=\frac{1}{r}+r \nonumber \\
		r=\frac{1}{2}(E-\sqrt{E+2}\sqrt{E-2}).
	\end{eqnarray}
	Plugging into the boundary conditions and manipulating gives us
	\begin{eqnarray}
		C_B^2=(E-r)(E-\mu_B).
	\end{eqnarray}
	This tells us that $r$ must be real. 
	Isolating for $r$, we get
	\begin{eqnarray}
		r_\pm=\frac{\mu_B\mp\sqrt{4C_B^2+\mu_B^2-4}}{2(1-C_B^2)}
	\end{eqnarray}
	and if $C_B=1$ then $r=1/\mu_B$. Since $r$ is real we require $4C_B^2+\mu_B^2>4$. The energy is then
	\begin{eqnarray}
		E_\pm&=&\frac{1}{r_\pm}+r_\pm \nonumber \\
		&=&\frac{(2-C_B^2) \mu_B \mp C_B^2 \sqrt{4C_B^2+\mu_B^2-4}}{2(1-C_B^2)}.
	\end{eqnarray}
	Then we calculate
	\begin{eqnarray}
		\label{v1}
		&&|v_1^\pm|^2=1-\nonumber \\ && \frac{C_B^2(\mu_B\mp\sqrt{4C_B^2+\mu_B^2-4})^2}{(1-C_B^2)(4(1-C_B^2)-(\mu_B\mp\sqrt{4C_B^2+\mu_B^2-4})^2)}. \nonumber \\
	\end{eqnarray}
	
	Now we look at the product
	\begin{eqnarray}
		r_+r_-=\frac{1}{1-C_B^2}.
	\end{eqnarray}
	If there are two solutions we must have $|r_+ r_-|<1$. This means that when $0<C_B<\sqrt{2}$, we have one solution whereas when $\sqrt{2}<C_B$, it is possible to have two solutions. 
	
	For $|r|<1$, we must have
	\begin{eqnarray}
		\label{singleBound}
		||\mu_B|-\sqrt{4C_B^2+\mu_B^2-4}|<2|C_B^2-1|
	\end{eqnarray}
	\begin{eqnarray}
		\label{doubleBound}
		||\mu_B|+\sqrt{4C_B^2+\mu_B^2-4}|<2|C_B^2-1|.
	\end{eqnarray}
	When there is only a single bound state, the only condition is \eqref{singleBound}. Then the solution for $|v_1|^2$ is given by
	\begin{eqnarray}
		\label{v1single}
		&&|v_1|^2=1-\nonumber \\&&\frac{C_B^2(|\mu_B|-\sqrt{4C_B^2+\mu_B^2-4})^2}{(1-C_B^2)(4(1-C_B^2)-(|\mu_B|-\sqrt{4C_B^2+\mu_B^2-4})^2)} \nonumber \\
	\end{eqnarray}
	As {$C_B$} $\rightarrow 0$, the above expression goes to $|v_1|^2 \rightarrow 1$ so long as there exists a bound state. Using \eqref{singleBound} in this limit of small $C_B$, the bound state exists if  $|\mu_B|>2$. This means for small $C_B$, there is a sharp transition where $|v_1|^2$ goes from 0 to 1, hence, the spectral switch concept.
	
	When a state starts in $|\psi(t=0)\rangle=\begin{pmatrix} 1 & 0 & 0 & 0 & \dots\end{pmatrix}^\intercal$, the long time expectation value $\langle n_0(t \to \infty)\rangle$ is $|v_1|^4$ in Eqs. \eqref{v1single} and \eqref{closedFormMain} of the main article.
	
	\bibliography{bibliographyNew}
\end{document}